\documentclass[aps,epsf,twocolumn,showpacs,superscriptaddress]{revtex4}
\usepackage{amsmath}
\usepackage{epsfig}

\begin{document}
\title{Advanced control with a Cooper-pair box: stimulated
       Raman adiabatic passage and Fock-state generation 
        in a nanomechanical resonator}

\author{Jens Siewert}
\affiliation{MATIS-INFM, Consiglio Nazionale delle Ricerche, and 
        Dipartimento di Metodologie Fisiche e Chimiche per l'Ingegneria,
 Universita di Catania,  I-95125 Catania, Italy}
\affiliation{Institut f{\"u}r Theoretische Physik, Universit{\"a}t Regensburg,
             D-93040 Regensburg, Germany}
\author{Tobias Brandes}
\affiliation{Department of Physics, The University of Manchester,
              Manchester, United Kingdom}
\author{Giuseppe Falci}
\affiliation{MATIS-INFM, Consiglio Nazionale delle Ricerche, and 
        Dipartimento di Metodologie Fisiche e Chimiche per l'Ingegneria,
 Universita di Catania,  I-95125 Catania, Italy}

\date{\today}

\begin{abstract}
The rapid experimental progress in the field of superconducting 
nanocircuits gives rise to an increasing quest for advanced 
quantum-control techniques for these macroscopically coherent
systems. Here we demonstrate theoretically that 
stimulated Raman adiabatic passage (STIRAP) should be possible 
with the quantronium setup of a Cooper-pair box. The scheme appears
to be robust against decoherence and
should be realizable even with the existing technology.
As an application we present a method
to generate single-phonon states of a nanomechnical resonator
by vacuum-stimulated adiabatic passage with the superconducting
nanocircuit coupled to the resonator. 
\end{abstract}


\maketitle

%%%%%%%%%%%%%%%%%%%%%%%%%%%%%%%%%%%%%%%%%%%%

%%%%%%%%%%%%%%%%%%%%%%%%%%%%%%%%%%%%%%%%%%%%
%%\section{Introduction}
%%%%%%%%%%%%%%%%%%%%%%%%%%%%%%%%%%%%%%%%%%%%

One of the most fascinating experimental breakthroughs of the 
recent past is the observation of quantum-coherent
dynamics in superconducting nanocircuits. It includes 
circuits exhibiting the dynamics of 
single `artificial atoms'~\cite{Nakamura99,Vion02,Chiorescu03}, two
coupled artificial atoms~\cite{Nakamura03,Majer05} and artificial atoms
coupled to electromagnetic resonators~\cite{Wallraff04,Chiorescu04}.
This development opens new perspectives to study quantum phenomena
in solid-state devices
that traditionally have been part of nuclear magnetic resonance,
quantum optics, and
cavity quantum electrodynamics. 
There exist already a large number of theoretical proposals for
such studies such as, e.g., the detection of 
geometric phases~\cite{Falci00}, 
the preparation of Schr\"odinger 
cat states in electrical and nanomechanical 
resonators~\cite{Marquardt01,Armour02}, 
cooling techniques~\cite{Martin04}, an analogue of
electromagnetically induced transparency~\cite{Orlando04}, and 
adiabatic passage in superconducting 
nanocircuits~\cite{Alec03,AdvSolSt04,Nori05}. 

One of the challenges is the preparation of Fock states
in a resonator coupled to a superconducting nanocircuit.
In quantum optics, the analogous problem has been solved 
both theoretically and experimentally~\cite{Parkins93,Henrich00}. 
The idea is to apply adiabatic passage to the dark state of a
three-level atom. Instead of driving the transition with two
classical fields as in conventional STIRAP~\cite{Bergmann98}, 
one of the external fields is replaced by
the quantum field of the cavity. % The consequence is that,
While the atom undergoes the transition, a single photon is emitted
into the cavity. 

In the following we will demonstrate the application of
this scheme to a Cooper-pair box operated as in the experiments
by Vion {\em et al.}~\cite{Vion02} (the so-called
quantronium device) coupled to a nanomechanical resonator. 
To this end, we need to make sure that adiabatic passage
in a three-level system using classical fields can be realized
with the quantronium setup of a Cooper-pair box. 
This circuit is appropriate for the substitution
of one of the classical driving fields by the quantum field
of the nanomechanical resonator without changing the functionality of the
Cooper-pair box. 
Coupling the resonator to the
nanocircuit and verification of the vacuum-assisted adiabatic
passage completes the analogue of 
the atom-cavity system in Refs.~\cite{Parkins93,Henrich00}.
We will discuss also the effects of decoherence on the scheme
in a real experiment.

We remark that, in principle, this program can be carried out 
for different regimes and setups of superconducting nanocircuits.
(An alternative realization is a flux qubit coupled to an electrical
resonator studied by Mariantoni {\em et al.}~\cite{Markus05}.)
We have chosen the quantronium as, on the one hand, it is very much 
analogous to the atom-laser system in quantum optics and, on the other
hand, it is a rather thoroughly studied system with respect to its
decoherence properties.

%%%%%%%%%%%%%%%%%%%%%%%%%%%%%%%%%%%%%%%%%%%%%%%%%%%%%%%%%%
%\section STIRAP & quantronium
%%%%%%%%%%%%%%%%%%%%%%%%%%%%%%%%%%%%%%%%%%%%%%%%%%%%%%%%%%

{\em Quantronium in a three-level STIRAP scheme.} Adiabatic
passage in three-level atoms is commonly realized with the
STIRAP technique which is based on a 
$\Lambda$ configuration of two hyperfine ground states
$|g\rangle$ and  $|u\rangle$ coupled to an excited state
$|e\rangle$ (with energies $E_{\mathrm{g}}$, $E_{\mathrm{u}}$, 
$E_{\mathrm{e}}$) by classical laser fields
$A_{\mathrm{g}} \cos{\omega_{\mathrm{g}}t}$, 
$A_{\mathrm{u}} \cos{\omega_{\mathrm{u}}t}$~\cite{Bergmann98,ScullyBook97}. 
In the frame rotating with the frequencies of the driving
fields $\omega_{\mathrm{g}}$, $\omega_{\mathrm{u}}$ the Hamiltonian reads
(applying the rotating-wave approximation)
\begin{equation}
 H_{\mathrm{rot.f.}}\  = \ \Delta|e\rangle\langle e| 
           + \frac{1}{2}(A_{\mathrm{u}}|e\rangle\langle u|
           +A_{\mathrm{g}}|e\rangle\langle g|+{\rm h.c.})\ \ 
\label{lambdaconfig}
\end{equation}
with the detuning $\Delta=E_{\mathrm{e}}-E_{\mathrm{g}}-\omega_{\mathrm{g}}=
                          E_{\mathrm{e}}-E_{\mathrm{u}}-\omega_{\mathrm{u}}$.
This Hamiltonian has a so-called dark state 
\begin{equation}
         |D\rangle\ =\ \frac{1}{\sqrt{|A_{\mathrm{u}}|^2 + |A_{\mathrm{g}}|^2}}
                 (A_{\mathrm{g}}\ |u\rangle\ -\ A_{\mathrm{u}}\ |g\rangle)\ \ .
\label{darkclass}
\end{equation}
From Eq.~(\ref{darkclass}) it can be seen that by slowly varying 
the coupling strengths $A_{\mathrm{u}}$, $A_{\mathrm{g}}$ the dark state can
be rotated in the two-dimensional subspace spanned by $|u\rangle$
and $|g\rangle$. 
For the so-called counterintuitive
scheme, the system is prepared in the state $|g\rangle$ with the
couplings $A_{\mathrm{g}}=0$ and $A_{\mathrm{u}}\neq 0$. 
By slowly switching $A_{\mathrm{u}}$ off while $A_{\mathrm{g}}$ is switched on,
the population can be transferred from state $|g\rangle$ to
state $|u\rangle$. Adiabaticity requires $|\dot{A_j}/A_j| < \omega_j$
($j=u,g$). 
\begin{figure}[h]
\resizebox{.48\textwidth}{!}{\includegraphics{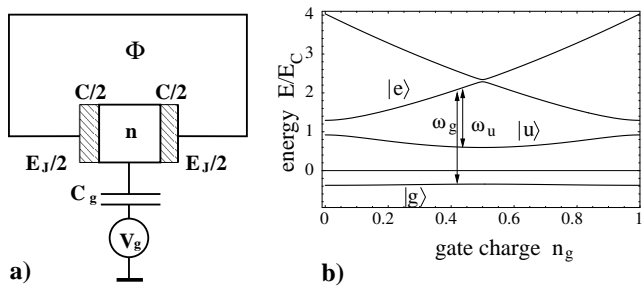}}
\vspace*{-7mm} \caption{\label{fig:setup} 
   a) In the quantronium setup, a superconducting island of
      total capacitance $C$ is coupled to a superconducting lead
      via two Josephson junctions. The gate voltage controls
      the offset charge of the island via the gate capacitance
      $C_{\mathrm{g}}\ll C$.
      The magnetic flux $\Phi$ represents
      another control parameter for the setup (here we choose $\Phi=0$).
   b) The lowest four energy levels of the quantronium 
      with $E_{\mathrm{C}}=E_{\mathrm{J}}$ as a function of gate charge.
      At the working point $n_{\mathrm{g0}}$ the three
      lowest levels can be used as a $\Lambda$ scheme 
      $|g\rangle$, $|u\rangle$, $|e\rangle$ with
      resonance frequencies $\hbar\omega_{\mathrm{g}}=
                     E_{\mathrm{e}}-E_{\mathrm{g}}$ and
      $\hbar\omega_{\mathrm{u}}=E_{\mathrm{e}}-E_{\mathrm{u}}$. 
      }
\end{figure}

In order to realize adiabatic population transfer with the
quantronium setup (see Fig.\ 1a) consider the corresponding 
Hamiltonian in the basis of the charge \mbox{states $|n\rangle$}
\begin{equation}
 H\ =\ \sum_n\ E_{\mathrm{C}} (n-n_g(t))^2|n\rangle\langle n| 
           + \frac{E_{\mathrm{J}}}{2} (|n\rangle\langle n+1| +\mathrm{h.c.})
\label{HamSaclay}
\end{equation}
with the charging energy $E_{\mathrm{C}}=(2e)^2/(2C)$ (where $C$ is the
total capacitance of the island and $(2e)$ the charge of a Cooper pair) 
and the Josephson coupling energy $E_{\mathrm{J}}$.
For the time being we assume $E_{\mathrm{C}}=E_{\mathrm{J}}$. The offset charge
$n_{\mathrm{g}}=C_{\mathrm{g}}V_{\mathrm{g}}/(2e)$ 
can be tuned with the gate voltage $V_{\mathrm{g}}$. In the quantronium
setup, the gate voltage (and hence the gate charge) has a d.c.\ 
part $n_{\mathrm{g0}}$ and an a.c.\ part
$n_{\mathrm{g}}^{\mathrm{ac}}=A\cos{\omega t}$ 
with a small amplitude $|A| \ll 1/2$. 

The STIRAP operation can be carried out between the three lowest energy
levels~(see Fig.\ 1b). For the working point $n_{\mathrm{g0}}$ values such as 
$n_{\mathrm{g0}}=1/2$ are
preferable that lead to low decoherence rates. 
However, at symmetry points small level spacings and
selection rules may impede the operation of the scheme~\cite{Nori05}. 
Therefore the working point needs to be chosen away from such
points, e.g., at $n_{\mathrm{g0}}=0.45$.
If two resonant frequencies $\omega_{\mathrm{g}}$, $\omega_{\mathrm{u}}$ 
are applied to the gate (see Fig.~1b), it is possible to adiabatically
transfer the population from the ground state $|g\rangle$ to the
first excited state $|u\rangle$. It is interesting to note that
the microwave field couples {\em diagonally} to the charge states (as opposed
to the dipole coupling in the three-level atom case). Nevertheless,
an effective Hamiltonian as in Eq.~(\ref{lambdaconfig}) is obtained
as only those off-diagonal matrix elements in the eigenbasis of
the driven Hamiltonian are important that couple two states 
resonantly~\cite{Kmetic86}.

\begin{figure}[bh]
 \resizebox{.48\textwidth}{!}{\includegraphics{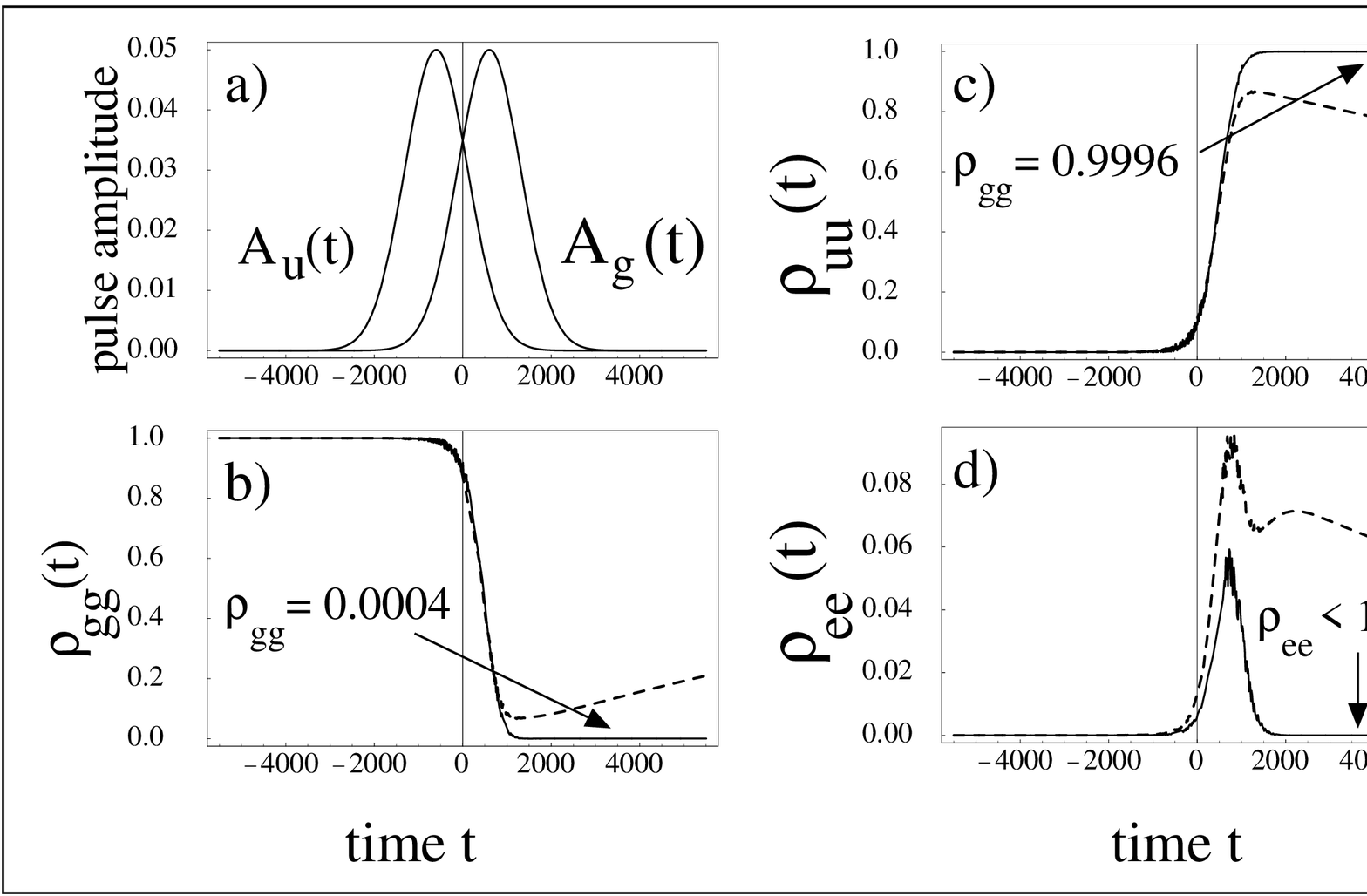}}
\vspace*{-3mm}
\caption{%\label{} 
   Population transfer by STIRAP in the quantronium setup
   ($n_{\mathrm{g0}}=0.45$). 
   a) Gaussian pulses are applied in the counterintuitive scheme.
   The maximum gate charge of the microwave fields are 
   $\max{(A_{\mathrm{u}}(t),A_{\mathrm{g}}(t))}=0.05$.
   For a charging energy of $E_\mathrm{C}=50$ $\mu$eV
   the time unit corresponds to about \mbox{$1.3 \times 10^{-11}$ s}.
   b)--d) Time evolution of the populations $\rho_{\mathrm{gg}}$,
               $\rho_{\mathrm{uu}}$, $\rho_{\mathrm{ee}}$
               without decoherence (solid lines) and with decoherence 
               (dashed lines). 
      The arrows denote the final populations in the ideal case
      (no decoherence).
      For the calculations with decoherence 
      we have used the decay rate $\gamma_{\mathrm{u}}=4.4\times 10^{-5}$
      and the dephasing rate 
      $\tilde{\gamma}=2.6\times 10^{-4}$. The latter rate corresponds
      to a dephasing time of about 50 ns.
      }
\end{figure}
In Fig.\ 2b--d (solid lines) we show the numerical solution of the Schr\"odinger
equation for the Hamiltonian Eq.~(\ref{HamSaclay}) with a gate
charge $n_{\mathrm{g}}(t)=n_{\mathrm{g0}}
                + A_{\mathrm{g}}(t)\cos{\omega_{\mathrm{g}}t}
                + A_{\mathrm{u}}(t)\cos{\omega_{\mathrm{u}}t}$ 
(zero detuning, $n_{\mathrm{g0}}=0.45$). 
Initially the system is prepared in the state 
$|g\rangle$. Then, two Gaussian-shaped microwave pulses are 
applied~(cf.\ Fig.\ 2a). We observe that a population
transfer to state $|u\rangle$ of nearly unit efficiency can be achieved.
The state $|e\rangle$ practically does not get populated during
the STIRAP procedure (cf.~Fig.\ 2d). 
Note that there are many parameters that may be used
to optimize the efficiency such as duration, delay, relative height
and over-all shape of the pulses, the detunings etc.~\cite{Bergmann98}.

%%%%%%%%%%%%%%%%%%%%%%%%%%%%%%%%%%%%%%%%%%%%%%%%%%%%%%%%%%
%\section{} {\em Effects of decoherence.} 
%%%%%%%%%%%%%%%%%%%%%%%%%%%%%%%%%%%%%%%%%%%%%%%%%%%%%%%%%%

{\em Effects of decoherence}. 
The functionality of solid-state
quantum-coherent devices is rather sensitive to various
(often device-dependent) sources of decoherence. 
In the quantronium, high-frequency noise that 
is mainly responsible for unwanted transitions, coexists 
with low-frequency noise which mainly affects calibration 
of the device and determines power-law reduction of the 
amplitude of the signal~\cite{Falci-prl,Ithier05}. 

A detailed analysis of decoherence in the STIRAP protocol
due to a solid-state environment is beyond the scope of this 
work.
Here we only estimate the feasibility of the protocol 
and argue that the main processes determining decoherence do not 
involve the level $|e\rangle$. These processes have been well 
characterized and, as a matter of fact, do not prevent 
very long decoherence times in the quantronium.  
We start our analysis from the quantum-optical master 
equation 
$    \dot{\rho} = \frac{i}{\hbar}[\rho,H^{\prime}] - \Gamma\rho $
where $\rho$ is the density matrix and $H^{\prime}$ is the 
Hamiltonian~(\ref{HamSaclay}) in the rotating frame~\cite{Kuhn99}. 
At low temperature
the dissipator $\Gamma \rho$ includes spontaneous decay rates
of the excited states $|e\rangle$, $|u\rangle$
as well as environment-assisted 
absorption between eigenstates in the presence of the
laser coupling. 
In quantum-optical systems the rate 
$\gamma_{\mathrm{u}\to \mathrm{g}}$ 
vanishes and the remaining processes mainly % would 
act towards depopulating states while they are not 
populated, and therefore hardly affect the protocol.
In contrast, 
STIRAP for the quantronium may be sensitive to the 
extra decay $|u\rangle\to |g\rangle$ 
involving the two low-lying states. 

An estimate of the 
effect of decoherence is achieved by studying the
master equation 
(written in the basis $\{|g\rangle,|u\rangle,|e\rangle\}$)
with the dissipator
\begin{equation}
\label{decoh}
(\Gamma\rho)_{ij} = \frac{\gamma_i+\gamma_j}{2}
                                          \rho_{ij}
         -(1-\delta_{ij})\tilde{\gamma}
         -\delta_{ij}\sum_k \rho_{kk}
                                     \gamma_{k\to i}
\end{equation}
where $\gamma_i=\sum_{k\neq i} \gamma_{\mathrm{i}\to \mathrm{k}}$.
The dissipator is taken time-independent (which 
overestimates decoherence) and includes all transitions
as well as a dephasing rate 
$\tilde{\gamma}$ accounting phenomenologically for low-frequency noise. 
For the decay rate of the second excited state we assume 
$\gamma_{\mathrm{e}}=\gamma_{\mathrm{e}\to\mathrm{u}}+ 
                     \gamma_{\mathrm{e}\to\mathrm{g}}=2\gamma_{\mathrm{u}}$. 
In order to obtain a realistic estimate of decoherence
effects, rates on the order of those observed in the experiments
of Ref.~\cite{Ithier05} are used.

The dashed lines in Fig.~2b--d show results 
for the solution of the master equation with the dissipator~(\ref{decoh}).
One recognizes immediately a remarkable robustness of the STIRAP
procedure against decoherence.
% As the level $|e\rangle$ hardly gets populated it appears reasonable
% that its decay does not hamper the procedure significantly. 
The main noticeable effects are the variation of
populations during the waiting time after finishing the pulse sequence
and a slightly increasing population of level $|e\rangle$.

Low-frequency 
noise is modeled more realistically as due to impurities which are static 
during each run of the protocol but may switch on a longer time scale, 
thus leading to statistically distributed level separations. 
Averaging determines defocusing of the signal. 
Fluctuations of $E_{\mathrm{e}}$ may be relatively large, 
but they represent {\em equal detunings} of both
microwave fields and do not affect STIRAP. 
On the other hand, fluctuations of the separation
between the two lowest eigenstates %% are small, 
are potentially detrimental since they determine fluctuations of the 
{\em difference of detunings}. This leads to a reduced 
efficiency of population transfer which, however, may be 
improved by optimizing the parameters of the protocol.

%
%
%

%%%%%%%%%%%%%%%%%%%%%%%%%%%%%%%%%%%%%%%%%%%%%%%%%%%%%%%%%%
%\section{} Coupling to a harmonic oscillator
%%%%%%%%%%%%%%%%%%%%%%%%%%%%%%%%%%%%%%%%%%%%%%%%%%%%%%%%%%

{\em Coupling the quantronium to a harmonic-oscillator mode.}
As we have demonstrated, STIRAP should be well within
reach of present-day technology for superconducting 
nanocircuits. Therefore one might hope to 
apply this technique similarly as in quantum optics 
for the preparation of peculiar quantum states. One
such application is the generation of Fock states in
a cavity coupled to a three-level atom~\cite{Parkins93}.
For this purpose, the Cooper-pair box
needs to be coupled to a harmonic oscillator
degree of freedom. 
The  generic coupling Hamiltonian is
$ H_{\mathrm{int}}= \lambda (a+a^{\dagger})(n-n_{\mathrm{g}}) $.
There are various ways to implement this Hamiltonian with
electrical resonator circuits~\cite{FalciPlastina03} and
transmission lines~\cite{Wallraff04}, or with nanomechanical
resonators~\cite{Armour02,Martin04}.
In the following we will explain that along these lines
it is possible to generate Fock states in a nanomechanical
oscillator.
\begin{figure}[bht]
\resizebox{.48\textwidth}{!}{\includegraphics{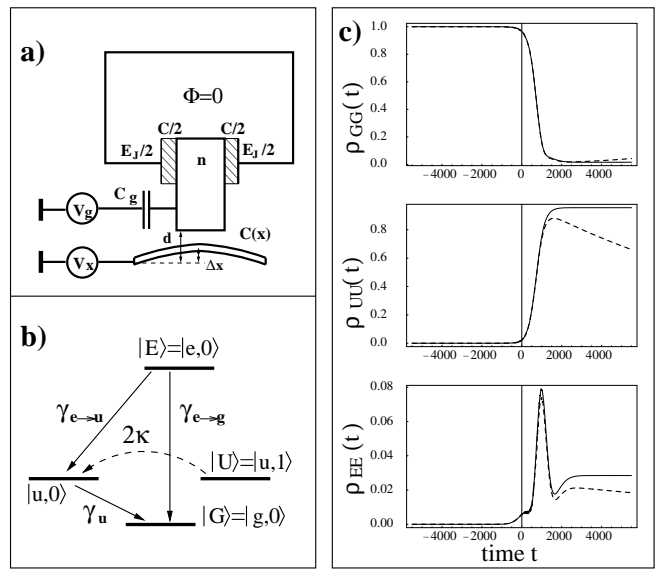}}
\caption{%\label{} 
   a) Coupled system of quantronium and nanomechanical osciallator.
   b) The four relevant states of the STIRAP scheme for Fock state
      generation in the presence of decoherence.
   c) Level population for the quantronium-resonator setup 
      in the presence of decoherence. Parameters are
      \mbox{$n_{\mathrm{g0}}(t) + n_{\mathrm{x}}(t)$}$=0.03$, 
     $\max{(A_{\mathrm{g}}(t),\lambda(t))}=0.05$,
      $\gamma_{\mathrm{u}}=4.4\times 10^{-5}$,
      $\tilde{\gamma}=2.6\times 10^{-4}$, $Q=5.0\times 10^{3}$.
      }
\end{figure}

The nanomechanical
oscillator (mass $m$) is coupled capacitively to the 
Cooper-pair box~\cite{Armour02,Martin04} via the 
position-dependent capacitance 
\mbox{$C(x)\simeq C_{\mathrm{x}} + 
 \Delta x (\mathrm{d}C(x)/\mathrm{d}x)$}, see Fig.~3a. 
Here $\Delta x$ denotes the oscillator 
displacement. The coupling can be tuned by the voltage $V_{\mathrm{x}}$.
Assuming $\Delta x\ll d$ and taking into account only a single
mode of the mechanical oscillator, the coupled quantronium--resonator system 
is described by the 
Hamiltonian~\cite{Armour02,Martin04}
\begin{eqnarray}
  H\ & = & \ H_{\mathrm{qu}} + H_{\mathrm{res}} + H_{\mathrm{int}}\ ,\ \ \
 H_{\mathrm{res}} =  \hbar \omega_{\mathrm{res}} a^{\dagger}a
\nonumber\\
  H_{\mathrm{qu}} & = &  \sum_n\ E_{\mathrm{C}} (n-n_{\mathrm{g}}(t)
                              -n_{\mathrm{x}}(t))^2|n\rangle\langle n| 
                           \ + \nonumber\\
     &&  \ \ \ \ \ +\ (E_{\mathrm{J}}/2) (|n\rangle\langle n+1| +\mathrm{h.c.})
%\nonumber
\\ 
  H_{\mathrm{int}} \ & =&  \  E_C\frac{2 C_{\mathrm{x}}V_{\mathrm{x}} }{2e}
                      \frac{1}{d}\sqrt{\frac{\hbar}{2m\omega_{\mathrm{res}}}} 
                      (a+a^{\dagger})(n-n_{\mathrm{g}})
\nonumber\\
                     & \equiv & \lambda (a+a^{\dagger})(n-n_{\mathrm{g}})
\nonumber
\label{mechcoupl}
\end{eqnarray}
where $d$ is the distance of the resonator from the island and
 $a^{\dagger}$, $a$ denote the creation and annihilation operators
for the nanomechanical oscillator. The total gate charge is now 
a sum of the box gate charge $n_{\mathrm{g}}(t)$ and 
          $n_{\mathrm{x}}(t)\equiv C_{\mathrm{x}}V_{\mathrm{x}}(t)/(2e)$
where $n_{\mathrm{g}}(t)=n_{\mathrm{g0}}(t)
                     + A_{\mathrm{g}}(t)\cos{\omega_{\mathrm{g}}t}$. 
The composed system is described
by the basis states $|j,N\rangle\equiv|j\rangle\otimes|N\rangle$
with the (uncoupled) quantronium eigenstates  $|j\rangle$ and 
the resonator Fock states $|N\rangle$. For the states relevant in
our discussion we will use the notation
$|G\rangle=|g,0\rangle$,
$|U\rangle=|u,1\rangle$, and 
$|E\rangle=|e,0\rangle$.

We assume that it is possible to prepare the vacuum state $|G\rangle$,
i.e., the oscillator frequency $\omega_{\mathrm{res}}$ 
has to be sufficiently large compared
to the temperature in the experiment (for a discussion of possible
values in an experiment see below). The population transfer is performed
from the initial state $|G\rangle$ via $|E\rangle$ to
the state $|U\rangle$. As the ``Stokes field'' $A_{\mathrm{u}}(t)$ is 
replaced by the vacuum
field of the cavity (which is coupled via the quantronium-resonator
coupling parameter $\lambda(t)$), a single phonon is emitted into the resonator
during the STIRAP operation. 
Again, the cavity field may trigger
transitions between eigenstates of the setup although it has only
terms diagonal in the charge basis due to mixing of charge states
by the Josephson coupling.

As mentioned above, $k_{\mathrm{B}} T < \hbar \omega_{\mathrm{res}}$ is
required. With typical temperatures of
$k_{\mathrm{B}} T \lesssim 30$ mK, oscillator frequencies
above \mbox{$\omega_{\mathrm{res}}> 2\pi \times 1$ GHz} are necessary
(which is at the limit of present-day technology~\cite{Roukes03}). 
Note also that the oscillator frequency needs to be resonant with
the quantronium transition $u\to e$. With a charging 
energy of $E_{\mathrm{C}}\approx 2E_{\mathrm{J}} \sim 35\ \mu$eV 
and $n_{\mathrm{g0}}\sim 0.03$ it is possible to have 
$\omega_{\mathrm{res}}\sim 2\pi\times 1.5$ GHz. With these parameters
one may hope to achieve similar decoherence effects as in the 
experiments of Refs.~\cite{Vion02,Ithier05} and, at the same time, to generate
the appropriate level spacings. 
Assuming the same decoherence rates as in the STIRAP process with
classical microwave 
fields (Fig.~2) and taking into account a finite quality factor
of the nanomechanical resonator $Q=\omega_{\mathrm{res}}/(2\kappa)=5.0\times
10^{3}$ we can numerically evaluate the time evolution of the coupled
system. Note that for this calculation it is necessary to take into account
also the state $|u,0\rangle$  which is not part of the
STIRAP scheme (see Fig.\ 3b) 
but contributes to reduce coherence of the population
transfer.  It can be seen that a highly efficient transfer
of the system to the state $|U\rangle$ should be feasible (cf.~Fig.~3c).

As to the detection, it would be desirable to directly measure the state
of the oscillator. However, it may be easier to
probe the state $|U\rangle$ via a %%time-resolved 
measurement of the quantronium
eigenstate. Either, one probes merely the final state $|u\rangle$.
Alternatively, the system can be viewed as a realization of the
Jaynes-Cummings model~\cite{ScullyBook97} and  one may 
try to detect Rabi oscillations 
between the states $|U\rangle$ and $|E\rangle$ induced by the
cavity field. To this end, 
the resonator-box coupling $\lambda(t)$ needs
to be set to the appropriate  value that facilitates the observation of
such Rabi oscillations (while $A_g\equiv 0$). Note that 
for this type of detection high-quality resonators
are required, and it is necessary 
to distinguish between the quantronium eigenstates $|u\rangle$ and
$|e\rangle$.

The procedure described here is not limited 
(at least in theory) to the generation of single-phonon states 
of the resonator~\cite{Parkins93}. 
The final state $|U\rangle=|u,1\rangle$ 
of the protocol described
so far may be changed (via a $\pi$ pulse in the quantronium with 
vanishing resonator-box coupling) into $|g,1\rangle$.
This state may serve as the initial state for another STIRAP transfer
$|g,1\rangle\rightarrow(|e,1\rangle)\rightarrow
|u,2\rangle$, etc. 

It is an important advantage of the STIRAP protocol for its realization
in solid-state devices that the efficiency does not depend sensitively 
on the absolute values of the couplings during the procedure. This makes
it robust against fluctuations in the environment.
Another advantage is its versatility. For example, instead of
changing the amplitudes of the driving fields it is possible 
to change the driving frequencies~\cite{Bergmann98}. This may be an 
option for a Cooper-pair box coupled to an electrical resonator such
as in Ref.~\cite{FalciPlastina03}
where it is easier to change the resonator frequency than the
capacitive coupling.
Interestingly, the protocol to generate Fock states 
can even be modified such that it suffices
to switch the couplings from $A_{\mathrm{g}}(t) \ll \lambda(t)$ at time $t$ 
to $A_{\mathrm{g}}(t^{\prime}) \gg \lambda(t^{\prime})$ at $t^{\prime}>t$ 
(see Ref.~\cite{Kuhn99}). That is, a single phonon (or photon) 
can be generated and emitted from the cavity with an `always-on'
cavity coupling. This may be interesting for setups where both 
coupling and resonator frequency are fixed such as 
in Ref.~\cite{Wallraff04}.

%%%%%%%%%%%%%%%%%%%%%%%%%%%%%%%%%%%%%%%%%%%%%%%%%%%%%%%%%%
%\section{acknowledgements}
%%%%%%%%%%%%%%%%%%%%%%%%%%%%%%%%%%%%%%%%%%%%%%%%%%%%%%%%%%

{\em Acknowledgments} This work has been supported 
financially by SFB 631 of the DFG.
JS would like to thank P.\ Schlagheck for pointing out to him 
Ref.~\cite{Henrich00} and D.\ Esteve for stimulating comments.
Illuminating discussions with A.\ Kuhn and M.\ Storcz
are gratefully acknowledged.

%%%%%%%%%%%%%%%%%%%%%%%%%%%%%%%%%%%%%%%%%%%%

\end{document}